\documentclass[acmsmall]{acmart}

\AtBeginDocument{%
  }
\setcopyright{none}
\copyrightyear{2026}
\acmYear{2026}
\acmJournal{TOCE}
\renewcommand\footnotetextcopyrightpermission[1]{}

\usepackage{booktabs}
\usepackage{multirow}
\usepackage{makecell}
\usepackage{array}
\usepackage{xcolor}
\usepackage{tikz}
\usepackage{pgfplots}
\usepackage{listings}
\usepackage{needspace}
\usetikzlibrary{arrows.meta,positioning,calc,shapes.geometric,patterns}
\pgfplotsset{compat=1.18}

\definecolor{accentA}{HTML}{1F77B4}
\definecolor{accentB}{HTML}{D95F02}
\definecolor{accentC}{HTML}{2CA02C}
\definecolor{softGray}{HTML}{F3F5F7}

\newcommand{\topone}{\textit{top-1}}
\newcommand{\topfive}{\textit{top-5}}
\newcommand{\topten}{\textit{top-10}}
\newcommand{\auc}{\textit{ROC--AUC}}
\newcommand{\gcj}{\emph{Google Code Jam}}
\newcommand{\gks}{\emph{Google Kick Start}}
\newcommand{\cj}{\emph{Code Jam}}
\newcommand{\ks}{\emph{Kick Start}}

\begin{document}

\title[Static and Process Evidence for Code Authorship]{Evaluating Static and Process Evidence for Code Authorship in Programming Education}

\author{Marek Horv\'ath}
\orcid{0009-0005-4649-2308}
\affiliation{%
  \institution{Department of Computers and Informatics, Technical University of Ko\v{s}ice}
  \city{Ko\v{s}ice}
  \country{Slovakia}
}
\email{marek.horvath@tuke.sk}

\renewcommand{\shortauthors}{Horvath}

\begin{abstract}
In programming courses, instructors may need to interpret whether a submission is consistent with a student's prior programming profile, especially when code similarity alone is inconclusive. Existing source-code authorship methods are often evaluated on programming-contest or open-source datasets, where reusable templates and local code patterns can produce strong author-related signal. Educational repositories present a different setting. Students solve shared assignments while their programming practices are still developing. This study uses task-aware evaluation to contrast these production contexts and tests whether repository-visible process features add information beyond final code in six matched educational comparisons. Contest data provide a high-signal contrast, with a \ks{} mean \topone{} of 0.938. Educational datasets produce substantially lower attribution performance. Adding process features raises the educational mean from 0.094 to 0.233 and mean pairwise verification \auc{} from 0.556 to 0.752. The comparisons show that measured signal depends on production context and that process patterns can complement weak final-code signal in educational repositories. Such models are therefore appropriate only as instructor-mediated decision support, not as independent proof of authorship.
\end{abstract}



\maketitle

\section{Introduction}

Source code is a formal artifact. It must be accepted by a compiler or interpreter, express an executable procedure, and often satisfy automated tests. These constraints might appear to leave little room for individual expression. Yet the same task can usually be implemented in several ways. Programmers choose how to decompose a solution, name variables and functions, use comments, organize control flow, reuse idioms, and revise a solution over time. Repeated choices of this kind form a programming style that can be observed, measured, and modeled.

Source code authorship analysis assumes that some part of this style remains stable across an author's work. That assumption is useful but conditional. Programming style is shaped by language, problem type, libraries, development environment, examples, templates, team conventions, and experience. Programmers can learn, imitate, adapt, or intentionally change their style. This instability is especially visible in education, where students are still acquiring programming habits and work under shared instructional constraints. Authorship analysis therefore cannot be reduced to selecting the classifier with the highest accuracy.

The literature on source code authorship has produced strong results in controlled settings, especially on programming-contest datasets such as \gcj{} and on other large code collections~\cite{caliskan-islam_-anonymizing_2015,abuhamad_code_2019,abuhamad_large-scale_2021,white_deep_2021}. These datasets are valuable because they contain many authors, many comparable tasks, and enough samples for training machine-learning models. They also have a particular character. Contest programmers often solve short algorithmic tasks and may reuse stable local templates, macros, type aliases, input-output routines, and other recurring fragments. Open-source repositories provide a different but also imperfect view because authorship can be affected by collaboration, project conventions, generated files, and copied code. Existing surveys and mapping studies repeatedly show that source code authorship research is dominated by closed-world attribution, static stylometric features, and a small number of benchmark datasets, while process and interaction evidence, verification scenarios, and realistic evaluation protocols remain less developed~\cite{neal_surveying_2018,kalgutkar_code_2020,he_authorship_2024,gurioli_stylometry_2024}.

Programming education changes the problem. Students in the same course may implement the same assignment, follow the same examples from lectures, start from the same skeleton, and submit code through the same infrastructure. The assignment can impose the input-output format, the expected algorithm, the public function signatures, or even parts of the directory structure. As a result, the strongest signal in a dataset may be the task rather than the author. A model can then appear successful while learning assignment-specific structure, not transferable author-specific behavior. This is a methodological problem as much as a modeling problem. If training and test data are split without respecting tasks, the reported accuracy can describe leakage from the assignment rather than authorship.

This article examines source code authorship in repositories created in programming courses. Programming education is the target application context. Contest datasets provide an external contrast for testing how production conditions affect measured signal. The central empirical comparison has two parts. The first is a task-aware contrast between contest and educational data. The second is a set of six matched educational comparisons that isolate the contribution of repository-visible process features beyond static code. TF-IDF feature documents, sequential models, verification, feature-removal analysis, and the remaining dataset families are used to test the scope, decision formulation, and source of the observed signal.

The empirical scope of the study is defined by four research questions.

\begin{itemize}
  \item \textbf{RQ1.} How does the strength of author-related signal differ between student programming repositories and programming-contest datasets?
  \item \textbf{RQ2.} When educational solutions are shaped by shared assignments, do repository-derived process features add useful information beyond the final source code?
  \item \textbf{RQ3.} How do evaluation protocol, authorship task, and representation shape the interpretation of results?
  \item \textbf{RQ4.} Which static and repository-derived process feature groups contribute to interpretable signal associated with student labels in course repositories?
\end{itemize}

The contributions follow the four research questions. A task-aware domain contrast quantifies the difference between educational and contest datasets for RQ1. Six matched static and static-process educational comparisons isolate the added value of repository-visible process variables for RQ2. RQ3 is addressed through complementary evaluation tasks and representations, and RQ4 through feature-removal and reduction analyses of one selected educational configuration. The combined evidence supports a bounded claim about context-dependent author-related signal, not a universal system for determining the author of arbitrary code.

\Needspace{12\baselineskip}
\section{Educational Motivation and Analytical Scope}
\label{sec:motivation}

Two needs shape the empirical design. The first is evaluating author-related signal under educational constraints. The second is determining whether repository history adds information when shared assignments constrain final code. Students learn, adopt examples, change tools, and adapt to tasks, so neither static nor process regularities are assumed to be immutable. Table~\ref{tab:motivation-directions} summarizes the resulting scope. Programming education is the target context, contest data provide a methodological contrast, and broader programmer-profile analysis remains contextual rather than a basis for hiring, surveillance, or productivity measurement.

\begin{table}[H]
\centering
\small
\caption{Motivating directions and how they shape the study.}
\label{tab:motivation-directions}
\begin{tabular}{>{\raggedright\arraybackslash}p{3.1cm}>{\raggedright\arraybackslash}p{4.8cm}>{\raggedright\arraybackslash}p{4.9cm}}
\toprule
\textbf{Direction} & \textbf{Practical question} & \textbf{Role in this paper} \\
\midrule
Academic integrity &
Is a submitted solution consistent with the student's previous programming profile? &
Motivates verification, attribution, task-aware evaluation, and conservative interpretation. \\
\midrule
Process-aware feedback &
What does the development history reveal beyond the final submitted program? &
Motivates process features derived from commits, timing, deadlines, and change volume. \\
\midrule
Programmer-profile analysis &
Which static and process traits appear stable enough to characterize a programming profile? &
Motivates feature-group analysis, but remains outside the direct application scope of the experiments. \\
\midrule
Methodological realism &
Do conclusions from contest-style datasets transfer to educational repositories? &
Motivates comparison between educational and contest data under controlled evaluation protocols. \\
\bottomrule
\end{tabular}
\end{table}

\subsection{Academic Integrity Beyond Code Similarity}

The most direct educational motivation is support for academic-integrity workflows in programming courses. Automated tests can show whether a submitted solution behaves correctly for a set of inputs. Code-similarity tools can show whether two submitted solutions resemble one another. Neither result answers a different question that instructors often face. The question is whether a solution is consistent with the student's usual way of solving programming tasks.

In this setting, code similarity and code authorship answer different questions. Similarity analysis compares artifacts. Authorship verification compares an artifact with a claimed profile, while authorship attribution asks which candidate label is most consistent with an unknown artifact. In a course setting, verification can be useful when there is no obvious copied counterpart inside the same submission set, or when the main issue is a sharp change relative to a student's earlier work. Attribution can also provide a ranking of plausible candidate students, although it is more restrictive and more sensitive to the available candidate pool. For practical educational use, a \topone{} attribution is the least defensible interpretation because it converts a relative ranking into a unique-label claim. Top-$k$ ranking and verification are closer to a review workflow. They can reduce the review space or rank specific profile-compatibility claims for human assessment. The operational objective is not to assign a label autonomously, but to prioritize a bounded set of cases or claims for contextual review.

For this reason, educational use should not amount to an automatic accusation. A functional solution can be atypical for legitimate reasons. The student may have learned a better approach, consulted course material, followed an example more closely, received permitted help, changed an editor, or solved a task with different structural demands. It can also be atypical because the final code was heavily influenced by an external source. Authorship analysis can surface cases that deserve interpretation, but it cannot replace that interpretation. In practical terms, a model can help prioritize cases such as the following.

\begin{itemize}
  \item submissions whose style differs sharply from earlier work by the same student,
  \item cases where pairwise similarity, student-profile evidence, and process evidence disagree,
  \item clusters of solutions with unusual structural or stylistic patterns,
  \item submissions that require instructor review rather than automatic judgment.
\end{itemize}

\subsection{Static Code and Repository Process Evidence}

Repository history can support understanding of how recorded work developed, not only what was submitted. Timing, continuity, change volume, and repeated revisions can prompt focused questions about decomposition, testing, or work organization that the final program cannot answer. These variables are neither direct measures of ability nor complete measurements of student behavior. They form an incomplete observational layer shaped by commit practice and course infrastructure. Their relevance is that repository-visible work patterns may retain variation when shared assignments compress differences in final code.

Static features describe the submitted artifact through size, layout, identifiers, comments, functions, control structures, operators, and language-specific constructs. Process features describe repository-visible timing, commits, deadline distance, and change volume. The first view asks what the program looks like; the second asks which regularities are visible in its recorded history. Their agreement or disagreement can inform review, but neither channel is inherently more trustworthy or meaningful outside the assignment and repository context.

\Needspace{10\baselineskip}
\subsection{Evaluation Scope and Claim Boundaries}

Contest and open-source datasets are useful but do not reproduce course conditions. Contest solutions may preserve reusable personal templates, whereas educational repositories contain common assignments, shared examples, novice programmers, and institutional constraints. In this study, the domain comparison tests whether measured signal depends on the production context, while task-aware evaluation tests whether it transfers beyond a specific assignment. The split protocol is part of the result, not an interchangeable implementation detail.

The practical stance follows from these motivations. Authorship models in education should be evaluated conservatively and used only as decision-support tools for prioritizing review and comparing static and process channels. Throughout the article, \emph{author-related signal} denotes patterns statistically associated with the available experimental label. \emph{Evidence} refers to a contextualized information channel considered alongside assignment context and human review, not independent proof of authorship, misconduct, originality, ability, or learning progress. The study does not evaluate adversarial imitation, deliberate obfuscation, pair programming or other collaborative authorship, or operational misconduct detection.

\section{Authorship Tasks, Evidence, and Related Work}
\label{sec:background}

The related literature is discussed through three connected choices. These are the authorship question, the representation, and the data-generating conditions. High accuracy can reflect strong author-related signal, but also a closed candidate set, repeated tasks across partitions, or reusable templates. The emphasis is therefore on the assumptions exposed by each experimental setting rather than on model chronology alone.

\subsection{Attribution and Verification as Distinct Tasks}

Most source-code authorship studies formulate the problem as closed-set attribution. An unknown program is assigned to one programmer from a known candidate set, commonly through multi-class classification. This formulation made it possible to demonstrate that lexical, layout, syntactic, and structural properties of code can carry sufficient information to discriminate among authors. The AST-based stylometric representation of Caliskan-Islam et al. established a widely used reference point, while later work scaled identification to larger candidate sets and introduced convolutional, recurrent, and language-oblivious representations~\cite{caliskan-islam_-anonymizing_2015,abuhamad_code_2019,abuhamad_large-scale_2021}.

Closed-set attribution is only one interpretation of programmer identification. Verification evaluates whether a program is consistent with a particular author, whether two programs are likely to share an author, or whether an artifact fits an existing author profile. A negative verification decision does not have to identify the true author. This changes both the practical meaning of the output and the experimental design. Attribution produces a ranking over candidates and can be evaluated using \topone{} and top-$k$ measures. Verification requires positive and negative comparisons, similarity scores or decision thresholds, and metrics such as \auc{}. In the educational experiments reported here, these author classes correspond to pseudonymous student labels rather than independently verified sole authorship.

This difference becomes especially important when the candidate set cannot be assumed to contain the true author. Metric-learning work has shown that learned distances can support both identification and verification, including set-based comparisons~\cite{white_deep_2021}. Verification has also been studied in binary and academic settings, where the objective is to accept or reject an authorship relationship rather than force every artifact into a known author class~\cite{ou_veribin_2024,romanov_integrated_2025}. Existing surveys nevertheless show a clear concentration on attribution, with verification receiving substantially less empirical attention~\cite{neal_surveying_2018,kalgutkar_code_2020,he_authorship_2024}. Reporting the two tasks separately avoids treating verification as attribution under a different label.

\subsection{Static, Sequential, and Repository-Derived Process Evidence}

The dominant source of evidence is the final code artifact. Explicit stylometric representations quantify recurring choices such as token and keyword frequencies, identifier use, line organization, comments, function decomposition, control structures, nesting, operators, or shapes derived from syntax trees. These features are attractive because they can be extracted without executing the program and can be inspected after a model has been trained. Classical classifiers, particularly random forests and support-vector models, therefore remain relevant even as learned representations have become more common.

Neural approaches shift part of feature design from manual extraction to representation learning. CNN and RNN models have been applied to token, character, TF-IDF, and sequential views of code, while metric-learning models construct an embedding space in which programs by the same author should be closer than programs by different authors~\cite{abuhamad_code_2019,abuhamad_large-scale_2021,white_deep_2021}. The literature does not imply that one model family is universally preferable. Explicit features offer a direct route to feature analysis and can work with smaller tabular datasets; sequential models preserve local ordering that aggregate counts discard, but require sufficient examples and may be harder to interpret. A model comparison is meaningful only when the representation and dataset are stated with it.

A measurable code property is not automatically intrinsic to its author. Formatting, time, dataset composition, language conventions, starter code, interfaces, tools, task difficulty, and programmer skill can all alter the observed profile~\cite{balla_code_2024,petrik_effect_2021,abazari_dataset_2023,caliskan-islam_-anonymizing_2015}. Such findings argue against interpreting a feature vector as a permanent personal signature.

Prior work has supplemented static stylometry with execution, interaction, or other artifact views~\cite{wang_integration_2018,romanov_integrated_2025}. This study instead uses repository-derived timing and change variables. They may preserve variation absent from short assignment-constrained programs, but remain proxies shaped by commit practice, course organization, collection infrastructure, and missing history. Static and process channels are therefore evaluated as complementary and imperfect rather than assumed to strengthen inference automatically.

\subsection{Educational Constraints and Sources of Confounding}

Programming competitions and public repositories dominate previous evaluation. \gcj{} enables large multilingual studies with many participant labels, but its compact solutions can preserve personal routines and templates~\cite{caliskan-islam_-anonymizing_2015,abuhamad_code_2019,abuhamad_large-scale_2021}. Open-source data introduce different ambiguity through collaboration, project conventions, generated or copied code, and authors outside the training distribution~\cite{gurioli_stylometry_2024}.

Educational repositories are neither a smaller \cj{} nor a cleaner public-repository benchmark. Shared specifications, languages, skeletons, and examples can increase task influence while developing programming habits reduce cross-task stability. Course repositories may also contain process information absent from final contest submissions. Educational verification has been studied~\cite{romanov_integrated_2025}, but remains less central than competition and open-source settings in existing surveys~\cite{neal_surveying_2018,kalgutkar_code_2020,he_authorship_2024}.

The main difficulty is not only dataset size, but the separation of three overlapping sources of variation.

\begin{itemize}
  \item \textbf{Author or student-profile variation:} repeated choices in code structure, naming, formatting, decomposition, and work routines.
  \item \textbf{Task variation:} the required algorithm, interface, language constructs, supplied materials, and expected project structure.
  \item \textbf{Process variation:} when and how a solution is developed, revised, committed, and submitted.
\end{itemize}

An evaluation split that allows the same task-specific structure to occur on both sides can make author variation appear stronger than it is. Conversely, a static-only analysis can underestimate author-related regularities when shared assignments constrain the final artifact. The educational setting therefore requires the representation and the split strategy to be considered together.

\subsection{Research Gap and Study Positioning}

The related literature shows that source code can contain measurable author-associated information, but it leaves several choices insufficiently connected. Attribution is usually evaluated more often than verification. Static stylometry is easier to obtain than process evidence. Contest datasets support scale but may reward stable templates, while educational repositories provide a more constrained and less frequently studied population. Standard random or cross-validation splits can also obscure whether a model recognizes authors or assignments. Finally, high predictive performance alone does not explain which feature groups produced the decision; explicit explanation techniques have been explored, but remain limited in the field~\cite{murenin_explaining_2020,hajihosseinkhani_authattlyzer-v2_2025}.

Table~\ref{tab:literature-positioning} links these observations to the design of the present experiments. No single representation is expected to resolve every limitation. The alternatives are instead placed within the same comparative framework so that differences between domains, tasks, and evidence sources remain visible.

\begin{table}[H]
\centering
\small
\caption{Relationship between recurring assumptions in related work and the design adopted in this study.}
\label{tab:literature-positioning}
\begin{tabular}{>{\raggedright\arraybackslash}p{3.0cm}>{\raggedright\arraybackslash}p{4.8cm}>{\raggedright\arraybackslash}p{5.0cm}}
\toprule
\textbf{Dimension} & \textbf{Recurring assumption or limitation} & \textbf{Design response in this study} \\
\midrule
Authorship task &
The true author is selected from a known candidate set. &
Attribution and verification are evaluated separately. \\
\midrule
Evidence &
The final source file is the principal source of author-related signal. &
Static features are compared with static-process representations and sequential code views. \\
\midrule
Data context &
Competition and public repository data dominate empirical evaluation. &
Educational repositories are contrasted with \cj{} and \ks{} datasets. \\
\midrule
Evaluation split &
Randomized partitions may retain problem-specific structure across training and testing. &
Task-aware splitting separates assignments where the data permit it. \\
\midrule
Interpretation &
Aggregate accuracy is emphasized more often than the origin of the signal. &
Top-$k$, \auc{}, model comparison, feature selection, and feature-group analysis are reported. \\
\bottomrule
\end{tabular}
\end{table}

\section{Study Datasets and Data Construction}
\label{sec:datasets}

The study is built on two source datasets that differ not only in scale but also in how a program becomes part of the dataset. The educational dataset was assembled from repositories used for coursework across several subjects and academic years. Its projects retain course and assignment context, and selected subsets also provide information derived from the development history. The contest dataset was reconstructed from public \gcj{} and \gks{} archives. It mainly represents selected submitted solutions to algorithmic tasks. These sources are not treated as interchangeable samples from one population. Their contrast is part of the experimental design.

\begin{table}[H]
\centering
\small
\caption{Processed source datasets used in the study.}
\label{tab:dataset-overview}
\begin{tabular}{>{\raggedright\arraybackslash}p{2.6cm}>{\raggedright\arraybackslash}p{2.8cm}rrr}
\toprule
\textbf{Source dataset} & \textbf{Main languages} & \textbf{Participant records} & \textbf{Projects} & \textbf{Files} \\
\midrule
Educational courses & C, Java & -- & 11,138 & 256,009 \\
\gcj{} & C/C++, Java, Python & 177,849 & 697,906 & 707,101 \\
\gks{} & C/C++, Java, Python & 110,876 & 397,232 & 397,232 \\
\bottomrule
\end{tabular}
\end{table}

In Table~\ref{tab:dataset-overview}, contest participant values are processed participant records summed across the corresponding partitions. The educational row is organized by projects and files. Participant-record counts are not reported at this level because candidate-label filters are applied only in the derived datasets.

\subsection{Educational Repositories}

The educational dataset consists of student projects from post-secondary programming courses. It includes C and Java assignments from the first semesters of study, object-oriented and component-oriented programming, systems-programming assignments, and a smaller advanced Java subset. It is therefore not a single-course benchmark. The included course subsets differ in language, project size, expected structure, and the stage at which students encountered the assignments. Table~\ref{tab:edu-summary} reports the retained data after cleaning, using anonymized descriptive labels rather than original course names.

\begin{table}[H]
\centering
\small
\caption{Educational dataset after cleaning. Course labels are anonymized descriptions rather than original course names.}
\label{tab:edu-summary}
\begin{tabular}{>{\raggedright\arraybackslash}p{3.3cm}>{\raggedright\arraybackslash}p{2.2cm}p{1.4cm}rr}
\toprule
\textbf{Anonymized course subset} & \textbf{Study stage} & \textbf{Language} & \textbf{Projects} & \textbf{Files} \\
\midrule
Introductory programming course A & 1st semester & C & 4,924 & 72,297 \\
Introductory programming course B & 2nd semester & C & 2,636 & 48,511 \\
Object-oriented programming course & 3rd semester & Java & 1,375 & 84,891 \\
Systems programming assignment sets & 3rd semester & C & 992 & 3,271 \\
Component-based programming course & 4th semester & Java & 1,076 & 44,254 \\
Advanced Java elective & master's level & Java & 135 & 2,785 \\
\midrule
\textbf{Total} & -- & -- & \textbf{11,138} & \textbf{256,009} \\
\bottomrule
\end{tabular}
\end{table}

The dataset was produced through a sequence of collection and normalization steps rather than through a single database export. Available projects were downloaded as source-tree archives and stored in a uniform local structure. The raw submissions varied in directory layout and contained configuration files, archives, temporary outputs, binary artifacts, IDE-generated directories, and other files unrelated to the intended analysis. Cleaning therefore traversed projects recursively, retained language-relevant files for each course subset, removed empty directories, and corrected structural irregularities introduced by archive export.

After cleaning, projects were reorganized around the relationship between a student label and that label's assignments. Original identifiers could occur in repository or directory names, so the experimental structure replaced them with identifiers of the form \texttt{<project-count>-id-<index>}. The project-count component preserves information needed to inspect data availability, while the index serves as an internal pseudonymous label. The mapping to original identifiers was used only to verify processing and is not part of the modeling data.

The student labels used in the educational experiments are experimental labels derived from course repository ownership or submission records. They are not independently verified ground truth of sole intellectual authorship. Permitted help, copied fragments, provided skeletons, generated material, and collaboration outside the recorded repository can affect the relationship between a label and the actual origin of a submitted program. The models therefore evaluate consistency with the available course label, not exclusive authorship.

The educational data were collected retrospectively from programming-course repositories hosted on an institutional Git infrastructure. Students submitted their assignments to this infrastructure as part of normal course participation, assessment, and course administration. The analysis used only pseudonymous labels and aggregate experimental outputs; detailed data-governance and ethics reporting is provided in the declarations after the conclusion.

Inclusion was decided at several levels. A retained file had to match the relevant language-oriented subset. A project had to contain at least one retained file and remain assignable to a course and task. A student label had to have at least one usable project. Requirements such as a minimum number of solutions per label were not imposed on the full dataset; they were introduced later when constructing particular experimental datasets.

\subsection{Contest Archives}

The contest dataset was derived from public archives of \gcj{} and \gks{} solutions. \gcj{} data cover 2008--2023, while the \ks{} partitions used in the dataset cover 2019--2022. The original material was distributed across compressed database archives whose organization and metadata differed between years. It could not be analyzed as a uniform directory of source files without reconstruction.

The processing pipeline identified available competition rounds, extracted attempt metadata and stored source content from the \texttt{raw\_data.sqlar} and \texttt{solutions.sqlar} databases, decoded and decompressed source files, and normalized task, language, and participant metadata. When several attempts by one participant were available for the same task, the last relevant scored attempt was retained. This reduced repeated near-duplicate versions and represented the final relevant state available in the archive.

File extensions and naming conventions were normalized across archive versions. In particular, Python variants such as \texttt{.py2}, \texttt{.py3}, and \texttt{.pypy} were mapped to \texttt{.py}. Older years required separate parsing because participant, task, and attempt information could be encoded directly in archive entry names. The resulting solutions were reorganized by competition partition, anonymized participant, and task. Contest identifiers use the same general pattern as the educational dataset, with the number of retained tasks replacing the project count. Public artifact records are listed in the data-availability statement for this version.

For the contest dataset, one project denotes one processed solution of one task by one participant. Project and file counts are equal in \ks{} and in the newer \cj{} partitions. They differ in the earlier \cj{} material because one archived solution can contain several files. This distinction explains why Table~\ref{tab:dataset-overview} reports 697,906 \cj{} projects but 707,101 files.

The reconstructed dataset is large, but it is not complete in a uniform historical sense. Some years contain partial source records or metadata without all corresponding programs, and archive schemas changed over time. Participant activity and task representation are also uneven. Many participants have few retained solutions, while a smaller group has substantially more. These properties are addressed when experimental subsets are sampled, but they remain limitations of the source dataset. The data also represent competition programming, where code is commonly written under time pressure and oriented toward algorithmic correctness and efficiency rather than toward the structure of a long-lived software project.

\Needspace{10\baselineskip}
\subsection{Comparative Role of the Source Datasets}

The two source datasets expose different forms of evidence and different sources of confounding. The educational dataset is smaller but preserves course and assignment context and, for selected subsets, process information derived from repository histories. The contest dataset supplies many participant labels and comparable algorithmic tasks, but primarily provides selected final submissions and competition metadata. Table~\ref{tab:corpus-comparison} summarizes the distinction that guides the later comparisons.

\begin{table}[H]
\centering
\small
\caption{Analytical roles and limitations of the two source datasets.}
\label{tab:corpus-comparison}
\begin{tabular}{>{\raggedright\arraybackslash}p{2.7cm}>{\raggedright\arraybackslash}p{5.0cm}>{\raggedright\arraybackslash}p{5.0cm}}
\toprule
\textbf{Dimension} & \textbf{Educational dataset} & \textbf{Contest dataset} \\
\midrule
Production setting &
Coursework completed during programming education. &
Algorithmic problems solved in competition rounds. \\
\midrule
Typical unit &
An assignment project that can contain multiple retained files. &
One processed participant solution for one competition task. \\
\midrule
Available context &
Course, assignment, pseudonymous student label, final files, and process records for selected subsets. &
Competition partition, task, participant, language, and selected submitted files. \\
\midrule
Principal dependency &
Shared specifications, examples, skeletons, and developing student practices. &
Competition format, task family, participant activity, and reusable local templates. \\
\midrule
Role in the study &
Primary setting for evaluating weak author-related signal and the contribution of process evidence. &
External contrast for evaluating methods on large collections of algorithmic solutions. \\
\bottomrule
\end{tabular}
\end{table}

The full source datasets are not the direct inputs to every reported experiment. Candidate-set sizes, minimum numbers of solutions, language restrictions, project-size groups, and the availability of process features are imposed only when the derived datasets are constructed. Keeping source-data construction separate from experimental sampling avoids presenting a scenario-specific filter as an inherent property of the underlying data.

\section{Experimental Design and Evaluation}
\label{sec:method}

The experimental design reflects a central property of the data. Authorship performance cannot be interpreted independently of the collection from which the samples were drawn. A single aggregate experiment would mix differences in candidate-set size, programming language, task structure, project scale, and the amount of prior work available for each label. The full source datasets were therefore used to derive smaller, controlled experimental datasets. Within this collection, the domain contrast and the matched static versus static-process educational datasets provide the primary empirical comparisons. The remaining families probe boundary conditions or provide diagnostic context rather than carrying equal argumentative weight.

\subsection{Derived Dataset Families}

The final experimental collection contains 38 derived datasets organized into ten families. A row represents one processed solution or project and retains a pseudonymous participant or student label, an assignment or task-group label, and the representation required by the corresponding experiment. Table~\ref{tab:dataset-families} summarizes the complete collection.

\begin{table}[H]
\centering
\small
\caption{Families of derived experimental datasets and the factor controlled by each family.}
\label{tab:dataset-families}
\begin{tabular}{>{\raggedright\arraybackslash}p{4.2cm}>{\raggedright\arraybackslash}p{7.8cm}}
\toprule
\textbf{Family} & \textbf{Controlled factor} \\
\midrule
Synthetic dataset & controlled synthetic clone scenario \\
Assignment type & open vs. closed educational assignments \\
Candidate set size & 100, 200, and 500 candidate labels \\
\ks{} & contest year and related contest transfer \\
Programming language & C, Java, and Python \\
LLM-period proxy & older vs. newer contest periods \\
Project size & small, medium, and large projects \\
Projects per profile & number of available solutions per label \\
Static-process educational data & static features plus repository-derived process variables \\
Static educational data & matching static-only educational datasets \\
\bottomrule
\end{tabular}
\end{table}

Table~\ref{tab:dataset-families} should be read hierarchically. The matched static and static-process educational datasets form the primary educational evidence, and the contest configurations provide the domain contrast. The supporting families address narrower questions. Candidate-set-size datasets examine growth from 100 to 200 and 500 labels. Projects-per-profile datasets vary the amount of prior evidence available for learning a profile. Language-specific datasets separate C, Java, and Python. Project-size datasets group solutions by file count. Assignment-type datasets compare educational tasks with different degrees of structural freedom.

The temporal family is deliberately described as an LLM-period proxy. It compares older and newer \cj{} years, but the archive does not record whether a participant used a generative model. Consequently, this comparison can reveal a change between time periods, not the use of an LLM in any individual solution. The synthetic dataset has a similarly bounded role. It is a controlled clone-generation scenario used as a reference point, rather than evidence about naturally produced student or contest code.

The six matched pairs provide the most direct comparison for the educational research questions. Each pair preserves the academic year and candidate-set size while changing whether repository-derived process variables are available. This design tests whether repository-visible work patterns add complementary author-related signal when shared assignments weaken the final-code signal. The workflow proceeds from dataset construction and cleaning to feature extraction, task-aware splitting, and attribution, verification, and feature-analysis outputs.

\subsection{Static, Process, and Sequential Representations}

The study uses two main views of a programming solution. The first is an explicit feature representation designed to retain interpretable properties of the final artifact and, where available, its development history. The second is a sequential representation that preserves the source code as a character or token sequence and allows a model to learn local patterns directly from the program text.

\paragraph{Static and repository-derived process vectors.}
Static features are extracted without executing the submitted program. For a project containing several retained source files, extraction is first performed at file level and then aggregated to one project-level vector. Counts are summed, average-valued metrics are averaged, and maxima are propagated as the maximum observed value within the project. The resulting groups describe code and line volume, layout, comments, identifier and variable use, functions and modularization, conditions and loops, operators and statements, and language-oriented elements such as preprocessor directives, imports, and libraries.

Process features are appended only for educational subsets with usable repository histories. They describe the recorded activity around the solution, including frequency and timing of commits, distance of work from the assignment deadline, time-of-day activity, number and size of changes, added and removed lines, changed files, and derived activity-intensity measures. These variables do not reconstruct every edit made by a student. They summarize activity visible in the repository and can therefore also reflect individual commit practices or work performed locally before a later commit.

The tabular preprocessing is fitted independently inside every experimental fold. Label and task identifiers are retained as metadata and excluded from the predictive feature matrix. Missing numeric values are replaced by the training-fold median and numeric columns are standardized. Missing categorical values are replaced by the most frequent training-fold value and categorical columns are one-hot encoded. The fitted transformation is then applied unchanged to the held-out fold. This ordering prevents information from the test assignment from influencing imputation, scaling, or category construction.

\paragraph{Sequential code representations.}
The sequential experiments provide a comparative representation analysis rather than a replacement for the explicit static-process representation used in the primary educational comparison. They operate on source code rather than on the aggregated feature table. Character input preserves whitespace, punctuation, comments, fragments of identifiers, and formatting choices. Token input applies lightweight lexical segmentation into identifiers, numbers, operators, and punctuation groups; it does not construct an abstract syntax tree or perform full language-specific parsing. The two representations expose different levels of evidence. Character sequences retain fine surface details and are comparatively language independent, whereas token sequences emphasize repeated lexical constructions and local programming idioms.

\subsection{Task-Aware Cross-Assignment Evaluation}

The main evaluation decision is to separate training and test data by assignment or task group. This decision is particularly important in the educational dataset, where students in the same course solve the same problem and may share an input-output contract, an expected algorithm, example code, or a project skeleton. A row-level random split can place these common structures in both training and testing and thereby reward recognition of the assignment rather than transfer of a student profile.

Let the dataset be
\begin{equation}
\mathcal{D} = \{(x_i, y_i, g_i)\}_{i=1}^{N},
\end{equation}
where $x_i$ is a representation of a solution, $y_i$ is the candidate label, and $g_i$ is an assignment or task group. The primary split is constructed so that test samples come from task groups not used in training. This is stricter than row-level random splitting and reduces the risk that the model learns the structure of a particular assignment instead of transferable label-associated signal. The protection is nevertheless incomplete. Different task groups can share a project skeleton, libraries, input-output conventions, examples, or closely related algorithms, while students or contest participants can reuse personal scaffolding across tasks. Holding out complete groups therefore controls direct task overlap but does not guarantee that label, course, and task-family effects are fully separated.

Operationally, the primary protocol follows a leave-one-group-out scheme. In each fold, one assignment or task group is held out and the remaining groups form the training data. All learned preprocessing steps, feature-document vocabularies, TF-IDF weights, label profiles, and model parameters are estimated from the training portion of that fold. Scores are then aggregated over the held-out groups.

A stratified random split is retained only as a diagnostic comparison. In that setting, labels with a single sample are removed because they cannot be represented in both partitions, and 30\% of the remaining samples form the test set. This baseline answers a different question from the primary protocol. It shows how well the model performs when solutions of the same or related tasks can occur on both sides of the split. It is neither an alternative main benchmark nor evidence of cross-task authorship identification, and it should not be used to rank the primary configurations.

\subsection{Attribution Models and Ranked Metrics}

Authorship attribution is modeled as multi-class classification over the available candidate labels. For a test sample $x_i$, the model produces scores over the candidate-label set $\mathcal{A}$ and ranks all labels. The primary metrics are \topone{}, \topfive{}, and \topten{}.

\begin{equation}
\mathrm{Top}\text{-}k =
\frac{1}{N}
\sum_{i=1}^{N}
\mathbf{1}\left(\operatorname{rank}_i(y_i) \leq k\right).
\end{equation}

\topone{} is ordinary accuracy. \topfive{} and \topten{} are also reported because, for a large candidate set, reducing the search to a short ranked list can be more informative than a forced single-label decision. Their reference level depends on the number of candidates.
\begin{equation}
\mathrm{RandomTop}\text{-}k =
\min\left(\frac{k}{|\mathcal{A}|}, 1\right).
\end{equation}
For this reason, the same \topten{} value does not have the same interpretation for 100 and 500 candidate labels.

\paragraph{Feature-based classifiers.}
The explicit-feature branch compares logistic regression, random forest, histogram gradient boosting, and nearest centroid classification. Logistic regression is allowed up to 4000 iterations. The random forest uses 400 trees, balanced class weights, and a fixed random state. Among these alternatives, the random forest was the most stable model over the heterogeneous combination of counts, proportions, timing variables, code-structure metrics, and correlated feature groups. It is therefore used as the principal feature-based model in the reported comparisons. A classical model also preserves a direct link between performance and the explicit feature groups used in the later analysis, consistent with prior code-stylometry studies that report strong tree-based baselines~\cite{tennyson_replicated_2013,kalgutkar_code_2020,hajihosseinkhani_authattlyzer-v2_2025}.

\paragraph{TF-IDF feature documents and score-level combination.}
The TF-IDF branch does not vectorize raw source code. Instead, it converts each row of the extracted feature table into a symbolic feature document. Categorical values become tokens of the form feature--value. Count variables are represented by repeated feature tokens, with repetition capped to prevent a large count from dominating the document. Continuous variables are assigned to quantile-based intervals and represented by interval tokens. TF-IDF weighting is then learned from the training documents and applied to the held-out assignment.

Linear SVM, ridge classification, and complement Naive Bayes are evaluated over this sparse representation. This branch tests whether recurring combinations of discretized feature values provide a useful alternative to direct tabular modeling. A further comparison combines the random-forest and linear-SVM outputs at score level. Their candidate-label scores are normalized and averaged with equal weight. The combination is intentionally late. The tabular and sparse representations remain separate, allowing the experiment to show whether their scores contain complementary information.

\paragraph{Sequential source-code models.}
The sequential experiments use character- and token-level convolutional models, recurrent models including GRU and LSTM variants, bidirectional LSTM models, and a transformer branch with a pretrained code encoder. One-dimensional convolutional filters target short recurring fragments, while recurrent models process the sequence in order and attempt to retain a wider context. The transformer tokenizer truncates or pads the code to its input length, and the sample representation is obtained from either the first special token or mean pooling over the encoder states included in the attention mask. A classifier then maps the resulting representation to the candidate labels.

All sequential families optimize the authorship-classification objective, use validation during training, and select the model state according to validation performance. They are evaluated on the 26 derived datasets for which the sequential experiment was performed. Sequential models are not assumed to be intrinsically more author-specific than explicit features. Local macros, input-output routines, formatting fragments, and identifier substrings can reflect an author, but they can also reflect a task or a reused contest template. Their results are therefore interpreted under the same task-aware constraints.

\subsection{Verification Methods and Metrics}

Verification addresses a narrower claim than attribution. Instead of selecting one identity from the complete candidate set, it asks whether a test solution is compatible with a specified profile. The experiments use the same training-side representations and task-aware folds, but convert them into scored label claims. A positive claim links a solution to its true label; a negative claim links it to another candidate.

\paragraph{Prototype verification.}
For each candidate label, the training vectors are averaged into a single profile. A test vector is compared with the claimed profile by cosine similarity. This creates a transparent reference method. A higher score means that the sample lies closer to the average representation for that label. Its limitation is equally direct. A single centroid compresses all observed variation in a student's or participant's work and may not represent several distinct task-dependent styles.

\paragraph{Pairwise verification.}
The learned verification method represents a sample--profile claim through the absolute difference between the sample vector and the claimed profile, augmented by their cosine similarity. Training data contain one positive claim for a sample and a limited number of negative claims involving other labels, avoiding construction of every possible negative pair. A class-balanced logistic regression then learns a compatibility score from these pair features. In contrast to direct cosine similarity, this model can assign different importance to individual dimensions of the sample--profile difference.

Both verification variants are evaluated with \auc{} and average precision (AP). \auc{} asks whether a randomly selected positive claim tends to rank above a negative claim without fixing a decision threshold. AP is stricter in the imbalanced candidate space because it summarizes precision across recall levels and is sensitive to the prevalence of positive claims. As the candidate set grows, each matching claim is evaluated among many non-matching claims, so AP can be numerically low even when AUC shows improved ranking. The two metrics are therefore complementary rather than contradictory. A threshold chosen through balanced accuracy is useful for diagnostic binary decisions, but it is not treated as a universal operating point across datasets because score distributions change with the candidate population and representation.

\Needspace{9\baselineskip}
\subsection{Feature Contribution and Representation Reduction}

Predictive performance alone does not show whether a model relies on process evidence, code layout, task-specific structure, or redundant measurements. The feature analysis therefore examines one selected educational static-process dataset in detail. It contains 651 records, 100 student labels, 7 assignment groups, and 88 features.

Individual features are assessed through several complementary criteria, including linear discriminative importance, regularized feature selection, random-forest importance, mutual information with the label, mutual information with the task, and statistical differences between labels. A task-adjusted information score is defined as
\begin{equation}
I_{\mathrm{label-task}}(f)
=
I(f;L)-I(f;T),
\end{equation}
where $f$ is a feature, $L$ the experimental label, and $T$ the task group. Positive values indicate that the feature carries more information about labels than about tasks; low or negative values warn that its apparent usefulness may be tied primarily to the assignment. For numeric variables, ANOVA and Kruskal--Wallis tests provide additional views of between-label differences. Categorical variables are examined with chi-square tests, with effect sizes expressed through $\eta^2$ for numeric and Cramer's $V$ for categorical features. The separate rankings are converted to rank-based scores and combined into a consensus ordering rather than relying on one importance measure.

The analysis also tests how much of the representation is necessary. Group-removal analysis measures the performance drop after removing process metrics, layout and line metrics, loops, functions, operators, conditions, variables, and other groups. Reduced subsets retain only the highest-ranked features and are evaluated with the same task-aware random-forest protocol used for the complete representation. Redundancy is examined through correlations among numeric features, with $|r|\geq0.85$ used to identify strongly correlated pairs, and through principal-component analysis after median imputation and scaling. These procedures distinguish three questions that a single importance ranking cannot answer. They show which groups contribute unique predictive value, which individual variables are strongly associated with labels rather than tasks, and how far the representation can be reduced while preserving attribution performance.

Throughout the experiments, each metric is tied to a particular dataset, candidate set, representation, and split. No attribution rank, verification score, or feature association is treated as independent proof of authorship. This constraint belongs to the method itself, not only to the later ethical discussion, because the experiments compare sources of author-related signal under controlled but imperfect observations.

\section{Results}
\label{sec:results}

The results are presented in the same order as the empirical argument. Family-level attribution establishes the production-context contrast, and the matched educational datasets provide the central test of complementary process signal. TF-IDF and sequential models place the representation choices in context, verification changes the decision formulation, and feature analysis examines which measurements contribute to the educational result.

\subsection{Attribution Varies Across Dataset Families}

The family overview uses assignment-separated evaluation over the 38 derived datasets. Table~\ref{tab:family-results} reports the arithmetic mean within each experimental family for the explicit-feature model and, in the last column, the corresponding mean \topone{} result for the TF-IDF feature-document model. These values provide context for the primary educational comparison; they do not give every family equal evidential status. Each row represents a different controlled comparison and should not be interpreted as if all families formed one homogeneous benchmark.

\begin{table}[H]
\centering
\small
\caption{Mean task-aware attribution results by dataset family. The TF-IDF column reports \topone{} accuracy.}
\label{tab:family-results}
\begin{tabular}{p{4.8cm}rrrr}
\toprule
\textbf{Dataset family} & \textbf{\topone} & \textbf{\topfive} & \textbf{\topten} & \makecell{\textbf{TF-IDF}\\\textbf{\topone}} \\
\midrule
\ks{} & 0.938 & 0.973 & 0.982 & 0.868 \\
LLM-period proxy & 0.909 & 0.969 & 0.980 & 0.830 \\
Candidate set size & 0.448 & 0.566 & 0.622 & 0.386 \\
Projects per profile & 0.349 & 0.466 & 0.524 & 0.285 \\
Static-process educational data & 0.233 & 0.456 & 0.580 & 0.149 \\
Synthetic data & 0.229 & 0.541 & 0.691 & 0.110 \\
Programming language & 0.110 & 0.208 & 0.282 & 0.073 \\
Static educational data & 0.094 & 0.237 & 0.337 & 0.059 \\
Project size & 0.073 & 0.168 & 0.242 & 0.044 \\
Assignment type & 0.004 & 0.014 & 0.027 & 0.005 \\
\bottomrule
\end{tabular}
\end{table}

Table~\ref{tab:family-results} is a broad overview, not the matched educational test. The static and static-process educational family means are descriptive summaries of heterogeneous years and candidate-set sizes. Their paired differences are examined separately in Table~\ref{tab:static-behavioral}. The synthetic row is a controlled reference scenario and is not evidence about naturally produced student or contest code.

The strongest family means occur in the contest data. \ks{} reaches \topone{} 0.938, and the LLM-period proxy family reaches 0.909. Their \topfive{} and \topten{} values are only moderately higher than \topone{}, indicating that the correct participant is frequently placed first rather than merely somewhere in a broad candidate list. These values demonstrate a strong repeatable signal in contest solutions under the evaluated protocol. They do not establish that programmers are universally identifiable outside this production context. The mixed candidate-set-size and projects-per-profile families occupy the middle of the ranking, with mean \topone{} values of 0.448 and 0.349, respectively.

The LLM-period value is a temporal comparison only. Because the archive contains no solution-level provenance for generative-tool use, neither this family mean nor the newer-period results reported below can identify LLM-assisted solutions or estimate the effect of LLM assistance.

The educational families show a different pattern. Static educational data average 0.094 at \topone{}, 0.237 at \topfive{}, and 0.337 at \topten{}. After process information is included, the corresponding family means rise to 0.233, 0.456, and 0.580. For candidate sets of 100 or 200 students under cross-assignment evaluation, a high first-rank label accuracy is not the expected operational outcome. The relevant result is that repository-visible process features consistently improve the ranking and place the correct student within a shorter candidate list more often. The project-size and assignment-type families remain weak under all reported top-$k$ measures. In particular, the assignment-type family reaches only 0.004 at \topone{} and 0.027 at \topten{}, showing little transferable author-related signal under this configuration.

TF-IDF follows the same ordering, but at a lower level for most families. Explicit features exceed TF-IDF by 0.070 in the \ks{} family, by 0.079 in the LLM-period family, and by 0.084 in the static-process educational family. Across all 38 datasets, the explicit-feature model has an average \topone{} advantage of 0.057 over TF-IDF. The only family-level reversal in Table~\ref{tab:family-results} is assignment type, where both representations are close to zero and TF-IDF is higher by 0.001.

\Needspace{34\baselineskip}
\subsection{Repository-Visible Process Evidence Improves Educational Ranking}

Because the family aggregates combine several academic years and candidate-set sizes, Table~\ref{tab:static-behavioral} gives the matched comparison for 2021, 2022, and 2024 with 100 and 200 candidate labels. The static and static-process datasets in each row describe the same year and candidate-set size; the difference is the inclusion of repository-derived process variables.

\begin{table}[H]
\centering
\small
\caption{Matched static and static-process educational datasets.}
\label{tab:static-behavioral}
\begin{tabular}{p{3.7cm}rrrr}
\toprule
\textbf{Year / candidate labels} & \makecell{\textbf{Static}\\\textbf{\topone}} & \makecell{\textbf{Static-}\\\textbf{process}\\\textbf{\topone}} & \textbf{Difference} & \makecell{\textbf{Static-}\\\textbf{process}\\\textbf{\topfive}} \\
\midrule
2021 / 100 labels & 0.138 & 0.276 & +0.138 & 0.527 \\
2021 / 200 labels & 0.086 & 0.222 & +0.136 & 0.401 \\
2022 / 100 labels & 0.111 & 0.292 & +0.181 & 0.534 \\
2022 / 200 labels & 0.077 & 0.175 & +0.098 & 0.376 \\
2024 / 100 labels & 0.084 & 0.250 & +0.166 & 0.502 \\
2024 / 200 labels & 0.068 & 0.181 & +0.113 & 0.399 \\
\bottomrule
\end{tabular}
\end{table}

\begin{figure}[H]
\centering
\begin{tikzpicture}
\begin{axis}[
  width=\linewidth,
  height=5.8cm,
  ybar,
  bar width=10pt,
  ymin=0,
  ymax=0.34,
  ylabel={Top-1},
  symbolic x coords={2021/100,2021/200,2022/100,2022/200,2024/100,2024/200},
  xtick=data,
  xlabel={Year / candidate labels},
  x tick label style={rotate=35,anchor=east},
  legend style={at={(0.5,1.02)},anchor=south,legend columns=2,draw=none,fill=none},
  ymajorgrids=true,
  enlarge x limits=0.08,
  grid style={draw=gray!20}
]
\addplot[fill=white, draw=black, postaction={pattern=north east lines, pattern color=black}] coordinates {
  (2021/100,0.138) (2021/200,0.086) (2022/100,0.111)
  (2022/200,0.077) (2024/100,0.084) (2024/200,0.068)
};
\addplot[fill=black!65, draw=black] coordinates {
  (2021/100,0.276) (2021/200,0.222) (2022/100,0.292)
  (2022/200,0.175) (2024/100,0.250) (2024/200,0.181)
};
\legend{Static, Static-process}
\end{axis}
\end{tikzpicture}
\caption{Repository-visible process features improve attribution ranking in matched educational datasets. Hatching and fill intensity distinguish static and static-process values without relying on color.}
\Description{A grouped bar chart comparing static and static-process top-1 attribution values for educational datasets from 2021, 2022, and 2024 with 100 and 200 candidate labels. Static bars use diagonal hatching and static-process bars use a darker solid fill. Static-process bars are higher in every configuration.}
\label{fig:static-behavioral}
\end{figure}
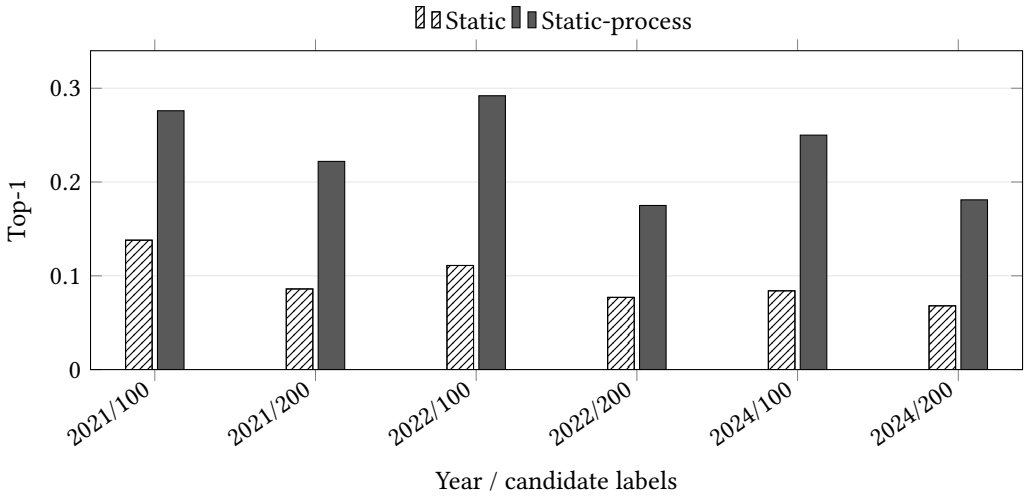

Table~\ref{tab:static-behavioral} and Figure~\ref{fig:static-behavioral} compare matched educational configurations. They are not a new general benchmark for authorship attribution. The absolute \topone{} values remain low enough that the result should be read as improved ranking support under cross-assignment conditions, not as reliable unique-label selection.

Repository-visible process features improve \topone{} in all six matched comparisons. The absolute gain ranges from 0.098 for the 2022 dataset with 200 labels to 0.181 for the corresponding 100-label dataset. Averaged over the six rows, the gain is approximately 0.139, which matches the difference between the family means in Table~\ref{tab:family-results}. The pattern is not driven by one exceptional year; it appears in every observed year and at both candidate-set sizes.

Candidate-set growth also has a consistent effect within the matched pairs. Moving from 100 to 200 labels lowers \topone{} for both representations in every year. For the static-process data, the decreases are 0.054 in 2021, 0.117 in 2022, and 0.069 in 2024. Despite this decline, the 200-label static-process configurations remain above their static-only counterparts. Their \topfive{} values range from 0.376 to 0.401, compared with 0.502 to 0.534 for the 100-label configurations.

Two points follow from these attribution results. Educational authorship remains substantially harder than contest authorship even after process information is added, but the process variables still provide a repeatable improvement within the educational setting. Their value lies in complementary ranking evidence under a demanding 100--200-candidate, cross-assignment protocol, not in supporting autonomous first-rank authorship decisions.

\Needspace{10\baselineskip}
\subsection{Explicit Features Outperform the TF-IDF Baseline}

Table~\ref{tab:representation-comparison} provides selected dataset-level comparisons for the explicit-feature model, the TF-IDF feature-document branch, their score-level combination, and the stratified random-split diagnostic. Unlike Table~\ref{tab:family-results}, these are individual configurations rather than family means.

\begin{table}[H]
\centering
\small
\caption{Selected \topone{} results for explicit features, TF-IDF feature documents, their score-level combination, and the random-split diagnostic.}
\label{tab:representation-comparison}
\begin{tabular}{p{5.1cm}rrrr}
\toprule
\textbf{Dataset} & \textbf{Features} & \textbf{TF-IDF} & \textbf{Combined} & \makecell{\textbf{Random}\\\textbf{split}} \\
\midrule
\ks{} 2021, 100 labels & 0.967 & 0.919 & 0.968 & 0.954 \\
\ks{} 2020, 200 labels & 0.948 & 0.866 & 0.948 & 0.934 \\
GCJ 2023, newer period & 0.937 & 0.868 & 0.937 & 0.893 \\
GCJ 2021, 500 labels & 0.752 & 0.624 & 0.753 & 0.703 \\
GCJ 2020, 5 projects/profile & 0.470 & 0.358 & 0.472 & 0.428 \\
Python & 0.248 & 0.172 & 0.250 & 0.201 \\
Educational 2022, static-process & 0.292 & 0.170 & 0.295 & 0.187 \\
Educational 2021, static-only & 0.138 & 0.084 & 0.140 & 0.082 \\
Educational data, 100 labels & 0.124 & 0.088 & 0.124 & 0.067 \\
Open assignments & 0.003 & 0.004 & 0.004 & 0.001 \\
\bottomrule
\end{tabular}
\end{table}

Table~\ref{tab:representation-comparison} is a diagnostic comparison of representations and split protocols in selected configurations. It is not an estimate of operational educational performance. Cross-row differences combine dataset conditions. The task-aware--random differences also reflect changed partition composition and size, so neither should be interpreted as a controlled domain effect.

The selected rows show the same broad ordering as the family means. Explicit features outperform TF-IDF in nine of the ten configurations. The exception is the near-zero open-assignment result. The largest selected difference is 0.122 for the 2022 static-process educational dataset. Score-level combination changes the feature-only result only slightly. Improvements in the displayed rows range from 0.000 to 0.003, and the average difference across all datasets is practically zero. TF-IDF therefore provides a comparison representation but does not add a systematic gain when its scores are averaged with the random-forest output.

The random-split column serves as a protocol diagnostic, not as an alternative benchmark or secondary leaderboard. In the selected rows it does not exceed the corresponding task-aware explicit-feature result. For example, educational 2022 static-process data reach 0.292 with assignment-separated evaluation and 0.187 in the reported random split, while GCJ 2021 with 500 labels reaches 0.752 and 0.703. These values should not be read as a controlled estimate of \emph{leakage gain}, because the random split changes the composition and size of the partitions in addition to mixing tasks. They show that the evaluation protocol materially changes the reported value and must be stated together with the result.

\Needspace{28\baselineskip}
\subsection{Character CNNs Lead the Sequential Comparison}

Sequential experiments cover 26 derived datasets. They compare what can be learned directly from code sequences and are not proposed as a replacement for the explicit static-process representation in educational use. The character CNN was the best-performing sequential model in 20 datasets, a token CNN in one, transformer models in three, and a token BiLSTM in two. Table~\ref{tab:sequential} reports selected configurations spanning contest, synthetic, language-specific, and educational data.

\begin{table}[H]
\centering
\small
\caption{Selected task-aware results of the best sequential model for each listed dataset.}
\label{tab:sequential}
\begin{tabular}{p{4.5cm}p{2.0cm}rrr}
\toprule
\textbf{Dataset} & \textbf{Model} & \textbf{\topone} & \textbf{\topfive} & \textbf{\topten} \\
\midrule
\ks{} 2021, 100 labels & char CNN & 0.990 & 0.993 & 0.996 \\
\ks{} 2020, 200 labels & char CNN & 0.982 & 0.991 & 0.995 \\
GCJ 2018, older period & char CNN & 0.971 & 0.989 & 0.991 \\
GCJ 2022, newer period & char CNN & 0.958 & 0.981 & 0.988 \\
GCJ 2023, newer period & char CNN & 0.957 & 0.986 & 0.992 \\
GCJ 2021, 500 labels & char CNN & 0.892 & 0.947 & 0.959 \\
GCJ 2020, 5 projects/profile & char CNN & 0.535 & 0.681 & 0.722 \\
Synthetic data & char CNN & 0.430 & 0.757 & 0.857 \\
Python & token CNN & 0.370 & 0.553 & 0.627 \\
Educational data, 100 labels & char CNN & 0.176 & 0.360 & 0.469 \\
Educational data, 500 labels & char CNN & 0.077 & 0.175 & 0.240 \\
Open assignments & token BiLSTM & 0.002 & 0.006 & 0.009 \\
\bottomrule
\end{tabular}
\end{table}

Table~\ref{tab:sequential} compares source-sequence representations under the same task-aware framing. It should not be interpreted as replacing the matched static-process educational comparison. The sequential branch does not use repository-visible process features and answers a representation question rather than the central process-evidence question.

The contest results remain high across different years and candidate-set sizes. The character CNN reaches 0.990 \topone{} for \ks{} 2021 with 100 labels and 0.982 for \ks{} 2020 with 200 labels. Even the \gcj{} configuration with 500 candidates reaches 0.892. The gap between \topone{} and \topten{} in these rows is small. It is 0.006 for \ks{} 2021 and 0.067 for \gcj{} with 500 labels.

Results outside the contest datasets are more varied. Synthetic data reach 0.430 at \topone{} but 0.857 at \topten{}, indicating that the sequential model narrows the candidate set more effectively than it makes a unique first-rank decision. Python is the only listed language-specific case won by the token CNN, with \topone{} 0.370. Educational data reach 0.176 with 100 labels and 0.077 with 500 labels, while open assignments remain near zero even with the best model in that dataset. The dominance of the character CNN across model families does not remove the domain gap observed with explicit features.

Longer training improved the average sequential results in the evaluated settings. For the character CNN, average \topone{} increased from 0.431 after five epochs to 0.458 after eight and 0.474 after twelve; the corresponding AUC values were 0.801, 0.824, and 0.830. For the transformer branch, increasing training from three to five epochs raised average \topone{} from 0.414 to 0.452 and AUC from 0.805 to 0.827. The improvement did not change the model-family ordering. Character CNNs remained the most frequent winners.

\subsection{Verification Reproduces the Domain Gap}

Verification views the same domain contrast from a different decision perspective. It asks whether a solution is compatible with a claimed student or participant profile. Table~\ref{tab:verification} reports selected pairwise and prototype results. For review prioritization, AUC indicates how consistently matching claims rank above non-matching claims. AP asks the stricter operational question of whether those ranked claims retain precision in a prevalence-imbalanced candidate space. Low AP can coexist with useful ranking improvement in AUC, while warning that the scores are not ready to serve as precise automated decisions.

\begin{table}[H]
\centering
\small
\caption{Selected pairwise and prototype verification results. P denotes pairwise verification; Pr denotes prototype verification.}
\label{tab:verification}
\begin{tabular}{p{4.5cm}rrrr}
\toprule
\textbf{Dataset} & \textbf{P-AUC} & \textbf{P-AP} & \textbf{Pr-AUC} & \textbf{Pr-AP} \\
\midrule
GCJ 2023, newer period & 0.987 & 0.429 & 0.963 & 0.564 \\
\ks{} 2021, 100 labels & 0.983 & 0.466 & 0.965 & 0.539 \\
\ks{} 2020, 200 labels & 0.979 & 0.290 & 0.953 & 0.516 \\
GCJ 2021, 200 labels & 0.970 & 0.153 & 0.933 & 0.374 \\
GCJ 2021, 500 labels & 0.959 & 0.067 & 0.924 & 0.220 \\
Educational 2024, static-process & 0.783 & 0.015 & 0.767 & 0.025 \\
Educational 2022, static-process & 0.760 & 0.024 & 0.793 & 0.067 \\
Educational 2022, static-only & 0.570 & 0.012 & 0.696 & 0.034 \\
Educational 2024, static-only & 0.513 & 0.005 & 0.631 & 0.012 \\
Open assignments & 0.492 & 0.001 & 0.554 & 0.002 \\
\bottomrule
\end{tabular}
\end{table}

Table~\ref{tab:verification} reports ranking evidence under controlled folds, not calibrated decision thresholds for course use. Verification is closer to an educational review workflow than closed-set attribution because it scores compatibility with a claimed profile.

Contest datasets again show the strongest separation between matching and non-matching claims. Pairwise AUC ranges from 0.959 to 0.987 in the displayed contest rows, and prototype AUC from 0.924 to 0.965. Increasing the candidate set affects AP more visibly than AUC. For example, the GCJ 2021 pairwise AUC remains 0.959 with 500 labels, but pairwise AP is 0.067 because the number of non-matching claims grows with the candidate space.

The matched educational comparisons reproduce the contribution of repository-derived process variables. For 2022, adding process features raises pairwise AUC from 0.570 to 0.760 and prototype AUC from 0.696 to 0.793. For 2024, the corresponding changes are from 0.513 to 0.783 and from 0.631 to 0.767. Across the educational families, mean pairwise AUC is 0.752 for static-process data and 0.556 for static-only data. AP remains low because of the large imbalance between matching and non-matching claims, but it also improves in each displayed matched comparison. Open assignments remain close to chance-level ranking in pairwise verification, with AUC 0.492.

\Needspace{10\baselineskip}
\subsection{Process Features Provide the Largest Contribution in the Analyzed Educational Dataset}

The detailed feature analysis uses one educational static-process dataset with 651 records, 100 student labels, 7 assignment groups, and 88 features. The full representation reaches 0.286 at \topone{}, 0.519 at \topfive{}, and 0.650 at \topten{}; chance-adjusted \topone{} is 0.278. This dataset is analyzed separately to determine which parts of that representation account for its predictive result.

Within this analyzed dataset, group-removal analysis identifies process and derived metrics as the most consequential group. Removing its 49 features lowers \topone{} from 0.286 to 0.163, an absolute loss of 0.123. Removing layout and code-line features produces the second-largest loss, 0.069. The remaining tested groups each change \topone{} by less than 0.02, as shown in Table~\ref{tab:feature-ablation}.

\begin{table}[H]
\centering
\small
\caption{Drop in \topone{} after removing feature groups from an educational static-process dataset.}
\label{tab:feature-ablation}
\begin{tabular}{p{4.7cm}rrr}
\toprule
\textbf{Removed group} & \textbf{Features} & \makecell{\textbf{\topone{} after}\\\textbf{removal}} & \makecell{\textbf{\topone{}}\\\textbf{drop}} \\
\midrule
Process and derived metrics & 49 & 0.163 & 0.123 \\
Layout and code lines & 16 & 0.217 & 0.069 \\
Loops & 5 & 0.267 & 0.018 \\
Functions & 3 & 0.272 & 0.014 \\
Operators and statements & 5 & 0.278 & 0.008 \\
Conditions & 3 & 0.283 & 0.003 \\
Variables & 4 & 0.284 & 0.002 \\
\bottomrule
\end{tabular}
\end{table}

Table~\ref{tab:feature-ablation} explains which feature groups contributed most in one selected educational static-process configuration. It does not imply that the same ranking will hold across courses, assignments, or repository policies.

The individual ranking is similarly concentrated on process information. The ten highest-ranked features are average, total, and maximum activity intensity; maximum distance of work from the deadline; average deleted lines; evening activity; average number of changes; average added lines; average distance of work from the deadline; and average function length. Nine of the ten describe repository activity or change history, while average function length is the only static code feature in this leading group.

Even so, the representation contains substantial redundancy. Table~\ref{tab:feature-reduction} reports selected reduced subsets. Thirty features retain 95.7\% of the full-representation \topone{} result, and 40 features retain 99.5\%. Removing highly correlated columns leaves 75 features and produces \topone{} 0.295, slightly above the unreduced value. Other tested subset sizes show the same non-monotonic pattern. The top 44 and top 60 features both reach 0.289, while the top 66 reach 0.296.

\begin{table}[H]
\centering
\small
\caption{Selected full and reduced feature representations for the feature-analysis dataset.}
\label{tab:feature-reduction}
\begin{tabular}{p{5.8cm}rrr}
\toprule
\textbf{Representation} & \makecell{\textbf{Variables}\\\textbf{retained}} & \textbf{\topone} & \makecell{\textbf{Relative}\\\textbf{\topone}} \\
\midrule
Full 88-feature representation & 88 & 0.286 & 1.000 \\
Top 26 ranked features & 26 & 0.260 & 0.910 \\
Top 30 ranked features & 30 & 0.273 & 0.957 \\
Top 40 ranked features & 40 & 0.284 & 0.995 \\
Top 66 ranked features & 66 & 0.296 & 1.038 \\
Representation after removing highly correlated features & 75 & 0.295 & 1.032 \\
\bottomrule
\end{tabular}
\end{table}

Table~\ref{tab:feature-reduction} asks whether the selected representation can be simplified without losing most of the observed ranking performance. The non-monotonic values caution against interpreting a shorter feature list as a stable causal explanation of authorship.

The dimensionality analysis is consistent with a representation whose information is distributed across several groups rather than concentrated in a handful of variables. Thirty-four principal components are required to explain 90\% of the numeric-feature variance and 43 components to explain 95\%. The correlation analysis identifies 29 strongly correlated feature pairs, and the largest correlated cluster contains six features. In the analyzed educational dataset, the feature-removal and reduction results show that process features provide the largest unique contribution, while a smaller but still multidimensional subset can preserve nearly all of the observed attribution performance.

Across the analyses, three patterns recur. Contest data support high attribution and verification values, including with large candidate sets. Educational results are lower but improve consistently when repository-derived process variables are added. Model complexity alone does not determine performance. Character CNNs are strong on contest sequences, whereas the explicit-feature representation remains essential in the educational setting because it can incorporate process variables and expose the contribution of individual feature groups.

\section{Discussion and Implications}
\label{sec:discussion}

The evidence does not support a general claim that source code either does or does not identify its author. Its central implication is narrower. In educational repositories, repository-visible process patterns contribute complementary author-related signal when shared assignments weaken the signal in final code. The contest comparison shows that signal strength depends on production context, while the remaining analyses clarify how representation, task structure, and decision formulation affect its interpretation. The section first synthesizes the four research questions and then discusses their combined methodological and educational implications.

\subsection{RQ1. Production Context Governs Signal Strength}

RQ1 concerns the strength of author-related signal in educational and contest code. The difference is substantial and appears in attribution, verification, and sequential modeling. Contest families achieve the highest attribution values. \ks{} averages 0.938 at \topone{}, while the LLM-period proxy family averages 0.909. Selected contest verification datasets reach pairwise AUC between 0.959 and 0.987. Character CNNs also retain high performance with large candidate sets, including \topone{} 0.892 for a \gcj{} dataset with 500 labels.

Educational results are consistently lower. Static educational datasets average 0.094 at \topone{}, and the assignment-type family remains close to zero. Even after repository-derived process variables are added, the static-process educational family reaches 0.233 rather than approaching the contest values. Sequential modeling shows the same pattern. The selected educational configuration reaches 0.176 for 100 labels and 0.077 for 500 labels, while open assignments remain near zero. The domain difference is therefore not specific to one classifier or one feature representation.

The most plausible interpretation is that the two source datasets expose different kinds and amounts of repeated author-related signal. Contest participants solve many short algorithmic problems and can reuse personal templates, macros, type aliases, input-output routines, formatting fragments, and identifier conventions. These elements are visible to both explicit features and character-level models. Educational solutions, in contrast, are produced under common assignment specifications, examples, skeletons, course conventions, and a narrower set of expected techniques. Students are also still developing their programming practices, so the style observed in one assignment may not transfer cleanly to another.

High contest accuracy should be read within this setting. It demonstrates a strong, repeatable author-related signal in contest solutions under the evaluated protocol; it does not show that the model has recovered a deep or immutable programmer identity. Persistent implementation routines, reusable templates, macros, type aliases, input-output code, and other stable contest idioms can all contribute to that signal. Task-aware evaluation tests transfer across problems, but recurring personal scaffolding can remain available across them. The result is valid for this production context without implying universal programmer identifiability.

The educational results matter for a different reason. Static \topone{} of 0.094 and static-process \topone{} of 0.233 can appear low in isolation, but they are obtained with 100--200 candidates, assignment-separated testing, common specifications, and developing student profiles. Under these conditions, the purpose is not to justify an autonomous unique-label decision. The substantive result is that repository-visible process features improve attribution ranking in every matched comparison and improve verification separation, even though the final-code signal remains weak. Contest benchmarks can therefore substantially overstate what is achievable for novice programmers completing common coursework.

\subsection{RQ2. Process Evidence Complements Final Code}

RQ2 asks whether repository-visible process features add information beyond final source code when educational solutions are shaped by shared assignments. The matched educational comparisons support this claim consistently. Adding repository-derived variables improves \topone{} in all six year and candidate-set configurations. The absolute gain ranges from 0.098 to 0.181 and averages approximately 0.139. The improvement is present for 100 and 200 labels and in all three observed academic years, so it is not attributable to one isolated subset.

Verification shows the same effect from a different angle. Mean pairwise AUC across static-process educational data is 0.752, compared with 0.556 for static-only data. In the 2022 matched comparison, pairwise AUC increases from 0.570 to 0.760 and prototype AUC from 0.696 to 0.793. In 2024, the corresponding changes are from 0.513 to 0.783 and from 0.631 to 0.767. The process representation improves not only selection among candidates but also compatibility scoring against a claimed student profile.

The selected-dataset feature analysis provides a local explanation consistent with the aggregate improvement. In that configuration, removing process and derived metrics causes the largest loss, reducing \topone{} from 0.286 to 0.163. Nine of the ten highest-ranked individual variables describe activity intensity, distance from the deadline, time-of-day activity, number of changes, or added and removed lines. These variables expose a layer absent from the submitted program. It captures the timing, rhythm, and scale of recorded repository activity.

Repository-derived process data should nevertheless be understood as a complementary decision-support channel, not as a replacement for static code or as direct measurements of behavior. The process group is strongest in the analyzed educational dataset, but layout and code-line features still produce the second-largest removal loss. The reduced-representation results also show that predictive information is distributed across multiple dimensions. Thirty to forty selected features preserve most of the full performance, yet the representation cannot be reduced to one timing variable, one commit count, or another isolated process marker.

Observed repository activity is not the same as the complete programming process. A large late commit can represent last-minute work, but it can also reflect local development followed by delayed synchronization. Frequent commits can reflect incremental work, repeated debugging, or a workflow encouraged by the course. Evening activity may be stable for one student but shaped by timetable constraints for another. The features are useful because their joint patterns differ across student labels; they do not carry a fixed pedagogical or ethical meaning on their own.

\Needspace{11\baselineskip}
\subsection{RQ3. Protocol and Representation Define the Result}

RQ3 concerns how evaluation protocol, authorship task, and representation shape the interpretation of results. None of these choices is merely technical; each defines a different question about authorship.

\paragraph{Task-aware versus random evaluation.}
The task-aware protocol asks whether a profile transfers to an assignment not used during training. This is stricter in concept than mixing solutions of the same task across partitions because it reduces direct access to assignment-specific structure. The random-split diagnostic produces different values, but the direction is not uniformly favorable. In the selected comparisons, random-split values are lower than the corresponding task-aware feature results. This does not invalidate the task-aware motivation. The two procedures also change fold composition, test size, label representation, and the number of training samples, so their numerical difference is not a pure measure of leakage. The result is that a performance value is inseparable from its split definition. Random and task-aware results answer different questions and should not be presented as interchangeable estimates of the same capability.

\paragraph{Attribution, ranked candidates, and verification.}
Attribution ranks a solution against the complete candidate set. This formulation is useful for studying how difficulty changes from 100 to 500 labels, but \topone{} alone is too restrictive for many educational cases. The static-process educational family averages 0.233 at \topone{} but 0.580 at \topten{}, showing that the representation can often narrow the candidate space even when it cannot select one student reliably. The practical value of such narrowing still depends on the original candidate count; \topten{} among 100 labels and among 500 labels has a different random reference level.

Verification asks whether a solution is compatible with a declared student or profile. This is closer to an educational scenario in which a submission is compared with a student's earlier work. AUC values show that educational representations contain useful ranking signal even where closed-set attribution is difficult. AP asks whether matching claims retain precision while being ranked among many non-matching claims. This prevalence-sensitive question is stricter. Low AP is therefore expected in the large imbalanced candidate space and does not contradict an AUC improvement, but it does rule out translating that improvement directly into operational precision. Verification is a better formulation for some use cases, not an automatic solution to class imbalance or uncertainty.

\paragraph{Explicit features and TF-IDF feature documents.}
Explicit features outperform the TF-IDF feature-document representation by an average \topone{} margin of 0.057 across the 38 datasets. Score-level combination produces practically no average improvement over explicit features alone. This indicates that the discretized TF-IDF documents mainly re-express information already present in the feature table rather than contributing an independent evidence source. The baseline remains useful because it demonstrates that direct modeling of heterogeneous numeric and categorical features is more effective than converting those same measurements into sparse symbolic documents.

\paragraph{Sequential models.}
Character CNNs are the most successful sequential family, winning 20 of 26 evaluated datasets. Their strongest results occur in contest code, where short local patterns can recur across solutions. Token CNN, BiLSTM, and transformer models win only a minority of datasets. Longer training improves average CNN and transformer results, but it does not change the overall ordering. Higher model capacity and broader context are therefore not sufficient when author-related signal is weak, heavily constrained by the task, or truncated in longer projects.

The most suitable representation depends on both domain and purpose. Character models are effective when reusable local fragments carry strong signal. Explicit representations are required when process evidence must be incorporated and when the purpose includes identifying which feature groups drive a decision. No single model family dominates every experimental role. The strongest predictive representation in contest data is not automatically the most informative or appropriate representation in educational data.

\subsection{RQ4. Signal in the Analyzed Educational Dataset Is Process-Dominated but Distributed}

RQ4 asks which feature groups contribute to interpretable signal associated with student labels in course repositories. In the analyzed educational static-process dataset, the strongest group is process and derived metrics. Its removal causes a \topone{} loss of 0.123, compared with 0.069 after removing layout and code-line features. Loops, functions, operators, conditions, and variables each produce smaller individual removal losses in that configuration.

At the individual-feature level, activity intensity dominates the ranking. Average, total, and maximum intensity occupy the first positions, followed by variables describing distance from the deadline, deleted and added lines, evening activity, and number of changes. Average function length is the highest-ranked static variable in the leading group. These results connect label-associated signal with three interpretable dimensions. They are work rhythm, temporal relation to the deadline, and scale of modification. The main static contribution concerns the amount and organization of code, particularly layout, line structure, and function length.

Interpretability does not imply that each feature is independently meaningful. Activity intensity is derived from repository records and can summarize several underlying actions. Added or deleted lines depend on project size and the type of assignment. Function length can reflect author preference, but also the decomposition naturally required by a task. This is why the task-adjusted feature score and group-removal experiments are more informative than a raw model-importance ranking alone. They ask whether a feature remains associated with authors after accounting for its relationship to the assignment and whether its removal changes predictive performance.

The reduction experiment further argues against treating authorship as a small set of fixed markers. Thirty features preserve 95.7\% of the full \topone{} result, and 40 preserve 99.5\%, but 34 principal components are still needed to explain 90\% of numeric variance. The representation contains 29 strongly correlated pairs, yet removing correlated variables does not damage performance. Author-related signal is therefore distributed and partly redundant. A compact profile is possible, but it still combines multiple process and code dimensions.

The relative importance of these groups is specific to the analyzed educational static-process dataset. In a course with mandatory auto-formatting, layout features could weaken. In an environment where repository use is inconsistent or optional, process variables could become noisier. In a larger software project, modularization and structural design might become more important than deadline-oriented activity. The feature ranking characterizes this configuration and suggests what future studies should measure; it is not a universal list of programmer characteristics.

\subsection{Authorship as Context-Dependent Consistency}

Across the four research questions, programmer authorship is better described as context-dependent consistency than as an immutable identity signature. A program records choices made by an author, but those choices are constrained by the task, language, examples, tools, and stage of learning. Contest data expose strong repeatable local and template-level patterns. Educational data expose a weaker final-code signal to which repository-visible process patterns add complementary author-related signal. Both are forms of consistency, but they do not have the same origin or the same expected stability.

This point matters when results are compared across benchmark families. A method that performs well on contest submissions may be exploiting precisely the properties that make contest datasets convenient. These include repeated short solutions, stable personal templates, and many samples per participant. Such performance is valid for that domain, but it does not establish transfer to novice code, collaborative software, or long-lived projects. Conversely, low educational performance should not be dismissed as an unsuccessful replication of contest authorship analysis. It reveals that the educational problem contains stronger confounding from assignments and less mature student profiles.

The temporal LLM-period family illustrates the same need for bounded interpretation. Newer and older contest years can be compared descriptively, and both periods contain strong author-related signal in the reported experiments. The dataset does not record whether generative tools were used, so the result cannot establish that LLM assistance strengthens, weakens, or preserves individual style. It only shows that the selected newer-period data remain highly attributable under the evaluated protocol. Any causal claim about generated code would require solution-level provenance that is absent here.

Style development also complicates the notion of a stable profile. A student may legitimately change naming, decomposition, testing habits, or repository use after learning a new technique. A mismatch with earlier work can therefore represent progress rather than misconduct. For longitudinal educational use, student profiles should be expected to evolve, and disagreement between recent and older work may itself be an object of analysis. The present experiments show cross-task regularities, but they do not define how quickly a profile should be updated or when style drift becomes educationally meaningful.

\subsection{Implications for Computing Education}

For computing education research, the results argue for studying authorship analysis as a contextual method for interpreting student programming work, not as a general detector. Programming courses combine repeated assignments, starter material, evolving student practices, and local repository rules; these conditions shape the measured signal as much as the classifier does. Studies in this area therefore need to report the course context, task split, label source, and process-data coverage alongside performance values.

For instructors, the practical role supported by these results is decision support, not automated authorship judgment. Static code, process history, pairwise similarity, assignment context, and instructor knowledge answer different questions. A useful workflow would treat them as separate evidence channels. An atypical static profile could trigger examination of the development history; unusual process activity could be checked against repository use and deadline circumstances; high similarity could be compared with whether the same patterns occur in prior work. Agreement among signals can prioritize a case, while disagreement is a reason for caution rather than a problem to hide.

Verification is particularly compatible with this workflow because it compares a submission with the declared student's profile instead of producing an unrestricted accusation against another candidate. Even then, the score should be accompanied by its reference distribution, the assignments used to build the profile, and interpretable differences in feature groups. A fixed threshold transported from one course or year to another would ignore the score shifts observed across datasets and candidate sets.

Repository-derived process variables also have a constructive use independent of misconduct detection. Timing, change volume, and activity continuity can support feedback about work organization. Repeated late concentrated changes may motivate advice to begin earlier or commit incrementally. Extensive revisions may identify a part of the task that caused difficulty. Group-level patterns can indicate that an assignment, example, or deadline structure affected many students. These interpretations remain hypotheses for discussion with the student or instructor; a repository trace does not by itself reveal motivation, effort, or understanding.

For future educational authorship studies, the minimum reporting unit should include the dataset origin, candidate-set size, number of solutions per label, task grouping, language, project size, process-data coverage, split protocol, and all reported metrics. Static and repository-derived process representations should be evaluated separately before they are combined. Attribution should be accompanied by ranked-candidate or verification results where appropriate, and feature-group analysis should identify whether the model relies on student-profile-related or task-related information. These practices would make results easier to compare and reduce the risk of treating a dataset-specific score as a general measure of programmer identifiability.

\section{Threats to Validity}
\label{sec:threats}

The results depend on the available labels, the recorded repository history, and the way tasks are separated during evaluation. The following limitations describe the main boundaries of the study.

\subsection{Labels and Process Proxies}

The educational labels come from course repository ownership or submission records, and contest labels come from participant submissions. They are necessary for supervised evaluation, but they are not verified records of sole intellectual authorship. Permitted help, copied fragments, starter code, generated material, or work outside the repository can affect the relation between a label and a submitted program. The experiments therefore measure consistency with the available label. They do not establish intent, originality, or exclusive authorship.

The static and process variables are also proxies. Static features can reflect personal habits, but they can also encode assignment interfaces, expected algorithms, examples, formatting tools, or templates. Process features describe commits, timing, deadline distance, and change volume visible in the repository. They do not reconstruct every edit. A student can work locally and commit late, combine many edits in one commit, or follow a workflow encouraged by the course. The feature values are useful as a joint profile, but no individual metric has a fixed meaning across students and assignments.

The same caution applies to the temporal contest comparison. The older--newer year grouping is an LLM-period proxy because the archive contains dates but no solution-level provenance for generative-tool use. Results from this family describe differences between selected historical periods. They cannot be used to infer whether an individual solution was generated, assisted, or unaffected by an LLM.

\subsection{Task, Dataset, and Transfer Limits}

Task-aware splitting is the main control against assignment leakage. Holding out complete task groups prevents the exact held-out assignment from appearing in the training portion of a fold, but it cannot remove every dependency between tasks. Different assignments can reuse the same project skeleton, libraries, input-output conventions, examples, or related algorithms. Contest participants can also reuse personal scaffolding across tasks.

Both source datasets required substantial reconstruction and filtering. The educational projects were exported from course repositories and cleaned by language-oriented rules. Process-data coverage is available only for selected subsets with usable histories. The contest dataset was reconstructed from public \cj{} and \ks{} archives whose schemas and source coverage changed over time. Some historical partitions contain incomplete source records, and participant activity is uneven. Experimental filters further determine which labels and projects are represented in each derived dataset.

The results should therefore not be generalized to every course, institution, language, or time period. Courses with pair programming, automatic formatting, different starter code, different commit policies, or other languages can produce a different balance between static and process signal. Profiles can also drift as students learn, change tools, or adjust their work habits.

\subsection{Metric and Educational-Use Limits}

The reported metrics answer different questions. \topone{} depends on the candidate set, while \topfive{} and \topten{} describe whether the review space can be narrowed. Verification \auc{} measures ranking of matching above non-matching claims. Average precision is stricter because positive claims are rare. AUC should not be translated directly into the precision of a deployed screening rule.

Feature importance has similar limits. Correlated variables can exchange importance, and a high association with a label does not show that a variable is stable across courses. The feature-removal and reduction analyses explain one selected educational dataset and support the matched comparisons. They are not a universal ranking of programmer characteristics.

Educational use should remain conservative. The practical safeguards are as follows.

\begin{itemize}
  \item do not infer ability, effort, intent, or misconduct from one score or one feature;
  \item interpret model output together with the assignment, repository history, permitted assistance, and other available evidence;
  \item preserve human review and allow the student to provide contextual explanation;
  \item use the result to prioritize review or support feedback, not as an automated judgment.
\end{itemize}

The appropriate role is review prioritization and contextual feedback, not automated judgment.

\Needspace{15\baselineskip}
\section{Conclusion}
\label{sec:conclusion}

This study examined source code authorship as a context-dependent problem rather than asking whether code can identify its author in general. Under task-aware evaluation, contest datasets retained strong, repeatable signal associated with local routines and reusable templates, whereas final code from shared educational assignments produced much weaker cross-assignment attribution. The clearest educational finding comes from the six matched comparisons. Adding repository-derived process variables raised mean \topone{} from 0.094 to 0.233 and mean pairwise verification \auc{} from 0.556 to 0.752. The detailed feature analysis supports this result within one selected configuration, where removing process and derived metrics caused the largest loss. It does not turn process measurements into direct behavioral measures or universal author markers. It shows that repository-visible work patterns can complement final-code evidence when the assignment constrains much of what students submit.

The broader implication is both methodological and practical. An authorship score reflects the source dataset, candidate set, task split, representation, and available experimental label, so high contest performance cannot be transferred automatically to educational repositories. At the same time, low educational \topone{} does not make the analysis irrelevant. Top-$k$ rankings and verification scores can still narrow a review space when they are interpreted with static code, repository history, assignment context, and a student's explanation. Educational use should therefore remain task-aware, review-oriented, and transparent. These methods can support prioritization, feedback, and further inquiry, but they cannot independently establish authorship, misconduct, originality, intent, or ability. Future validation should test transfer across institutions and time, examine changing student profiles, and calibrate verification under realistic class imbalance before any operational use is considered.

\section*{Acknowledgments and Declarations}

\subsection*{Data Availability}

Public archival release is planned only for non-student artifacts, including the cleaned programming-contest dataset, the synthetic C mutation dataset, and the literature-extraction and bibliographic artifact. Public artifact records are listed below. Student course repositories, original repository names, identity mappings, and raw repository histories are not publicly released because they contain coursework submissions, development histories, and potentially identifiable work patterns. The article reports aggregate dataset summaries, derived dataset definitions, and evaluation results. Any release of derived educational feature tables or scripts that depend on local course paths, identity mappings, or raw student histories would require separate institutional approval; public artifact scripts that do not require student repositories can be linked in a non-anonymous artifact statement.

\begin{anonsuppress}
\paragraph{Public artifacts.}
Electronic Appendix 2 is the cleaned \gcj{} and \gks{} programming-contest dataset for source code authorship analysis, covering 2008--2023. It is available as Zenodo record 18901762, DOI: \url{https://doi.org/10.5281/zenodo.18901762}. Electronic Appendix 3 is the Synthetic C Code Mutation Dataset, available as Zenodo record 20312253, DOI: \url{https://doi.org/10.5281/zenodo.20312253}. Electronic Appendix 4 contains the literature-extraction script and curated bibliographic dataset for programmer-attribution research, available as Zenodo record 18641274, DOI: \url{https://doi.org/10.5281/zenodo.18641274}.
\end{anonsuppress}

\subsection*{Ethics and Student-Data Statement}

The educational data were collected retrospectively from programming-course repositories hosted on an institutional Git infrastructure. Students submitted assignments to this infrastructure as part of normal course participation, assessment, and course administration. The research team had authorized access to the repositories as members of the teaching team for course administration and assessment. Before analysis, direct identifiers and original repository names were removed, and experiments used pseudonymous student labels. The identity mapping was used only for data-processing verification and was not included in modeling data.

Raw repositories, development histories, and identity mappings are not publicly released because they contain student coursework and potentially identifying work patterns. The reported results are aggregated and were not used for decisions about individual students. No additional intervention, survey, or data collection from students was conducted for this study. No formal ethics-board review was obtained for this retrospective analysis. We report this explicitly and limit the study to pseudonymized data, restricted raw-data access, and aggregate experimental outputs.

\subsection*{Use of AI Tools}

ChatGPT and Grammarly were used to support language editing and the preparation of LaTeX tables and figures. They were not used to generate experimental data, derive labels, run experiments, or create reported results. All reported data and numerical results were obtained from the described repository-processing and experimental programs.

\bibliographystyle{ACM-Reference-Format}
\bibliography{toce-references}

\end{document}